\documentclass[
  aps,
  pra,
  reprint
]{revtex4-1}

\usepackage[english]{babel} 						
\usepackage[latin1]{inputenc} 					
\usepackage{times} 										
\usepackage{graphicx}
\usepackage{amsmath}										
\usepackage{amssymb}							
\usepackage{bbm}

\begin{document}


\newcommand{\ket}[1]{|#1\rangle}
\newcommand{\bra}[1]{\langle #1|}
\newcommand{\ketvac}{\ket{0}}		
\newcommand{\ketone}{\ket{1}}
\newcommand{\bravac}{\bra{0}}	
\newcommand{\braone}{\bra{1}}	

\newcommand{\drho}{\hat{\rho}}
\newcommand{\anni}{\hat{a}}
\newcommand{\crea}{\hat{a}^\dag}
\newcommand{\po}{\hat{p}}
\newcommand{\qo}{\hat{q}}
\newcommand{\xo}{\hat{x}}

\newcommand{\inlab}{{\text{in}}}
\newcommand{\outlab}{{\text{out}}}
\newcommand{\reflab}{{\text{ref}}}
\newcommand{\flslab}{{\text{false}}}

\newcommand{\Win}{W^{\inlab}}
\newcommand{\Wout}{W^{\outlab}}
\newcommand{\Wref}{W^{\reflab}}

\newcommand{\Chi}{{\text{\Large$\chi$}}}
\newcommand{\N}{\mathcal{N}}
\newcommand{\derv}[1]{\partial_{#1}}

\numberwithin{equation}{section}
\renewcommand{\theequation}{\arabic{section}.\arabic{equation}}


\title{Quantum teleportation of nonclassical wave-packets: An effective multimode theory}

\author{Hugo Benichi}
\email{hugo.benichi@m4x.org}
\affiliation{Department of Applied Physics, The University of Tokyo, Tokyo, Japan}

\author{Shuntaro Takeda}
\affiliation{Department of Applied Physics, The University of Tokyo, Tokyo, Japan}

\author{Noriyuki Lee}
\affiliation{Department of Applied Physics, The University of Tokyo, Tokyo, Japan}

\author{Akira Furusawa}
\affiliation{Department of Applied Physics, The University of Tokyo, Tokyo, Japan}

\date{\today}

\begin{abstract}

We develop a simple and efficient theoretical model to understand the quantum properties of broadband continuous variable quantum teleportation.
We show that, if stated properly, the problem of multimode teleportation can be simplified to teleportation of a single effective mode that describes the input state temporal characteristic.
Using that model, we show how the finite bandwidth of squeezing and external noise in the classical channel affect the output teleported quantum field.
We choose an approach that is especially relevant for the case of non-Gaussian non-classical quantum states and we finally back-test our model with recent experimental results.

\end{abstract}

\pacs{42.50.Dv, 42.50.Ex, 42.65.Yj, 03.65.Ud}

\keywords{teleportation, continuous variable, quantum information, photon subtraction}

\maketitle


\section{Introduction}
\label{section_intro}

As a striking example of quantum communication protocol, teleportation was discovered early on in the development of the field of quantum information processing.
With either qubit \cite{Bennet93} and continuous variable flavors \cite{Vaidman94}, experiments were soon to follow \cite{Zeilinger97,Furusawa98}.
Until now continuous variable teleportation has only been performed with the class of so-called Gaussian states \cite{Furusawa98,Yoshikawa07,Yukawa08}.
However, this alone is not sufficient for universal quantum computation where non-Gaussianity of some kind has been shown to be necessary \cite{Braunstein99}.
Although non-Gaussian nonclassical states of light that would allow for such universal operations have been available experimentally for some time in the continuous variable regime \cite{Ourjoumtsev06,Jonas06,Wakui07}, the major challenge of actually manipulating these states in some Gaussian protocol context beyond simple generation has remained mostly unaddressed.

Some recent experimental work has reported on successful continuous variable teleportation of a strongly nonclassical state of light \cite{Lee11}. 
In this experiment, a close approximation of a Schroedinger's cat state generated with the photon-subtraction protocol \cite{Dakna97} is sent through a continuous variable teleporter.
The quality of teleportation is high enough that the output teleported state is also a nonclassical state with a negative Wigner function.
The use of a non-Gaussian nonclassical state as an input state is the first most noticeable feature of this experiment.
Although Gaussian states teleportation has been amply studied, due to the complex nature of non-Gaussian states and especially mixed non-Gaussian states, only few general results exist for this case.
Maybe the most general condition for successful teleportation of non-Gaussian non-classical states is the necessary but not sufficient $2/3$ threshold on fidelity \cite{Ban04}.
In \cite{Mista10}, the theoretical work closest to the experimental conditions of \cite{Lee11}, teleportation success is investigated for the case of a mixture of vacuum $\ketvac$ and one photon $\ketone$ as an input state.
On top of these difficulties, to accommodate with the transient nature of the input state used, the teleporter used in \cite{Lee11} operates on a broad range of frequencies.
This is the second most noticeable feature of this experiment in contrast with typical continuous variable experiments, which only manipulate narrow sidebands of light.
To our knowledge, there are actually very few results relevant to the case of multimode teleportation.
In \cite{Noh09}, multimode teleportation of a quantum field is investigated from the point of view of temporal fluctuations using the photon correlations function $g^{(2)}(\tau)$.
In \cite{Loock00}, it is shown how to adapt the single-mode teleportation protocol of \cite{KimBrauns98} to teleportation of a multi-mode field with finite teleportation bandwidth.
Both these works attack the problem of multi-mode teleportation from the Henseinberg picture and additional considerations are required to handle the case of non-Gaussian nonclassical input states.

Our main objective in this paper is to attempt to answer both these issues with a theoretical model as simple and efficient as possible.
In Sec. \ref{section_basics}, we first briefly introduce the teleportation protocol with its usual phase-space formulation and define a criterion of success.
In Sec. \ref{section_input}, we describe a realistic model of a nonclassical non-Gaussian state that faithfully models the input states of \cite{Lee11}.
With this model, we are able to predict the success of teleportation in a way similar to \cite{Mista10}.
In Sec. \ref{section_multi}, we use the Heisenberg picture to approach multimode teleportation as teleportation of a quantum field.
We then show how to reduce this quantum field to a single effective mode that describes the temporal properties of the input state.
In Sec. \ref{section_noise}, we show how to take into account any external noise spectrum in the broadband teleportation operation.
Finally in Sec. \ref{section_conclu}, we compare our model with the recent experimental results of \cite{Lee11} and conclude.


\section{Basics}
\label{section_basics}

Deciding on success of continuous variable teleportation is a non-trivial problem as it is closely related to the kind of input states and entanglement used, as well as the specific protocol or quantum circuit teleportation is actually used for.
For the Gaussian case, the fidelity $F = \langle\psi_{in}|\hat{\rho}_{out}|\psi_{in}\rangle$  is the usual figure of merit, though $F$ loses much of its meaning as a benchmark figure when more general non-Gaussian mixed states are used as input states.
While Gaussian states can be fully characterized by their first and second moments, which allow figures like fidelity to have some general and useful meaning, such an approach fails with non-Gaussian states.
Because non-classicality itself is an ambiguous property for continuous variable systems with infinite-dimensional Hilbert spaces, it is even more complex to decide on a relevant success criterion for continuous variable teleportation of non-classical quantum states.
In this paper we consider the input $\Win$ and output $\Wout$ Wigner functions of a teleportation process.
We adopt as a criterion of success the successful transfer of negative features of the Wigner function.
Provided $\Win$ is itself a negative Wigner function and having for $\Win$ a precise algebraic expression including the relevant experimental parameters, we want to know what are the requirements on these parameters and on the teleportation process for successful retrieval of negativity in $\Wout$.
Furthermore, we restrict ourselves to the Braunstein-Kimble scheme described in \cite{KimBrauns98} where the teleportation can be expressed in phase space as the following convolution 
\begin{equation}
  \label{phasespace_tele}
  \Wout = \Win \circ G_{e^{-r}},
\end{equation}
with $r$ the Einstein-Podolsky-Rosen (EPR) correlation parameter and $G_\alpha(q,p)$ a normalized Gaussian of standard deviation $\alpha$ ($\hbar = 1$).
In this case, teleportation of nonclassical features such as negativity has been shown to require 3 dB of squeezing \cite{Ban04}, or equivalently a vacuum fidelity of $F \geq 2/3$, which is also called the no-cloning limit \cite{Grosshans01}. 
Precisely speaking, 3-dB is a lower bound for unity gain teleportation of negativity of any pure or mixed state. 
Recent work has shown that, given the precise shape of the input state and amount of anticorrelation in the teleportation quantum channel, there actually exists strategies to surpass the 3 dB threshold by tuning the gain of teleportation \cite{Mista10}.
However, unless extremely pure entanglement is used, typical experimental antisqueezing imposes virtually unity gain operation. 
Furthermore, as this tuning is input dependent, the teleportation setup losses its universal characteristic.

The 3-dB threshold is only a lower bound to negativity teleportation and we would like to have a model that predicts better the success or failure of negativity teleportation.
In the general case, this is too broad a problem to handle as teleportation is known to be input dependent.
From now on, we focus our analysis on the specific case of the photon-subtracted squeezed vacuum that was used as an input state in \cite{Lee11}.
This family of quantum states has recently attracted a lot of interest both experimentally \cite{Ourjoumtsev06,Jonas06,Wakui07} and theoretically \cite{Dakna97,Molmer06,Sasaki06}.
From an optical parametric oscillator (OPO) producing a squeezed vacuum $ \hat{S}_{s}\ketvac$ with squeezing parameter $s$ called the signal mode, a fraction $R$ of the output called the trigger mode is tapped and sent to a photon resolving detector to herald non-Gaussian states.
Various single-mode and multi-mode models exist for this protocol \cite{Dakna97,Molmer06,Sasaki06}, and they are essentially equivalent in the limit of small $s$ and $R$.
To start, we will assume that a detection event projects $\hat{S}_{s}\ketvac$ on the photon-subtracted squeezed vacuum $\anni \hat{S}_{s}\ketvac$ equal to a squeezed photon $\hat{S}_{s}\ketone$.
The Wigner function $\Wref$ of this reference state is written 
\begin{equation}
\Wref(q,p) = 2(e^{2s}q^2+e^{-2s}p^2-1/2)G_{1/\sqrt{2}}(e^{s}q,e^{-s}p),
\label{input_ref}
\end{equation}
and has a maximal central negativity of $\Wref(0,0) = -1/\pi$.
Although this is a specificity of this particular input state, from now on we will use the value of the Wigner function at the origin of phase space as the figure of merit for negativity teleportation.
Applying teleportation equation \eqref{phasespace_tele} on Eq. \eqref{input_ref}, we find for output negativity $W_{\text{ref out}}(0,0)$
\begin{equation}
W_{\text{ref out}}(0,0) = \frac{(2e^{-2r}+1)(2e^{-2r}-1)}{\pi\left((1+2e^{-2r})+8e^{-2r}\text{sh}^2(s)\right)^{3/2}},
\label{ref_neg_threshold}
\end{equation}
which indeed yields $W_{\text{ref out}}(0,0) \leq 0$ for $r \geq \ln\sqrt{2}$ as expected. 
$W_{\text{ref out}}(0,0)$ will become negative only if the parameter $r$ is greater than $\ln \sqrt{2}$, equivalent to 3 dB of squeezing.


\section{Realistic input state}
\label{section_input}

Equation \eqref{ref_neg_threshold} is, of course, of little interest since an actual experimental input state will virtually be a mixed state and a more realistic model of input is required.  
The experimental input state used in \cite{Lee11} happens to fit well a simple loss model where the experimental Wigner function $\Win$ can be modeled from $\Wref$, the reference state, by applying "beam-splitter losses" $1-\eta$ equivalent in phase space to the operation:
\begin{equation}
  \Win(x,p) = \frac{1}{\eta}\left(\Wref \circ G_\lambda\right)\left(\frac{x}{\sqrt{\eta}},\frac{p}{\sqrt{\eta}}\right),
  \label{phasespace_loss_on_ref}
\end{equation}
with $\lambda = \sqrt{\frac{1-\eta}{2\eta}}$ \cite{Leonhardt}.
The phase space transformation \eqref{phasespace_loss_on_ref} is derived from the action of a fictitious beam splitter with transmission coefficient $\eta$, which transforms input coherent states $\ket{\alpha}$ into $\ket{\eta\alpha}$.
As was shown in \cite{Agarwal07}, it is also possible to express the action of this beamsplitter with a master equation acting on the density matrix $\drho$.
In this case, we would obtain
\begin{equation}
\frac{d}{dt}\drho(t) = L[\drho(t)],\quad L[\drho] = [\drho\anni,\crea] - [\drho\crea,\anni].
\label{master_1}
\end{equation}

\begin{figure}[!h]
  \includegraphics[height=\columnwidth,angle=-90]{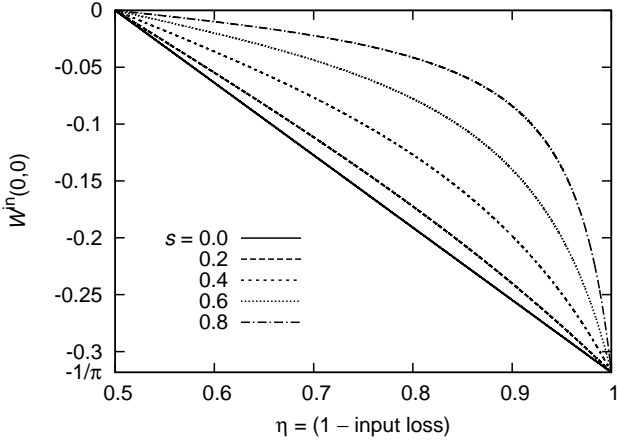}
  \caption{\label{fig_input_neg}Input Wigner function negativity $\Win(0,0)$ as a function of $\eta$ for different values of the squeezing parameter $s$.}
\end{figure}

\begin{figure}[!h]
  \includegraphics[height=\columnwidth,angle=-90]{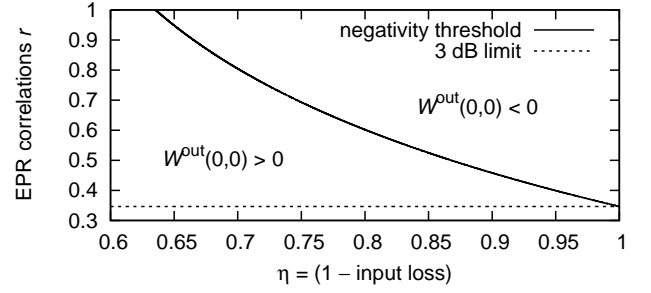}
  \caption{\label{fig_neg_thre}Negativity threshold as a function of $r$ and $\eta$.}
\end{figure}

By using the previous algebraic expression \eqref{input_ref} of $\Wref$ together with the transformation \eqref{phasespace_loss_on_ref}, we can obtain the exact expression of $\Win$ with any mathematical software.
We rather are interested in the central negativity given by
\begin{equation}
  \Win(0,0) = (1-2\eta)/\pi\left(1+4\eta(1-\eta)\text{sh}^2(s)\right)^{3/2},
  \label{phasespace_neg_input}
\end{equation}
where the negativity threshold $\Win(0,0) = 0$ depends only on $\eta$: $\eta\geq 0.5$ implies $\Win(0,0) \leq 0$ (see Fig. \ref{fig_input_neg}). 
Using this model of imperfect input state we investigate negativity teleportation of $\Win$ by concatenating Eqs. \eqref{phasespace_loss_on_ref} and \eqref{phasespace_tele}.
The two successive Gaussian convolutions are reduced to one, while the phase space rescaling survives the teleportation. Eventually $\Wout$ happens to be written in the same form as $\Win$,
\begin{equation}
  \Wout(x,p) = \frac{1}{\eta}\left(\Wref \circ G_{\lambda'}\right)\left(\frac{x}{\sqrt{\eta}},\frac{p}{\sqrt{\eta}}\right),
  \label{phasespace_tele_out}
\end{equation}
where $\lambda$ has been changed to $\lambda' = \sqrt{ \lambda^2 + \left( e^{-r}\right)^2 / \eta }$ in a way similar to classical amplifiers input/output SNR rules.
We remark that besides the degradation of the input negativity $\Win(0,0)$, the loss parameter $1-\eta$ has also the effect of decreasing the effective correlation parameter $r$ to $r' = r + \ln\sqrt{\eta} < r$. 
\begin{figure}[!h]
  \includegraphics[height=\columnwidth,angle=-90]{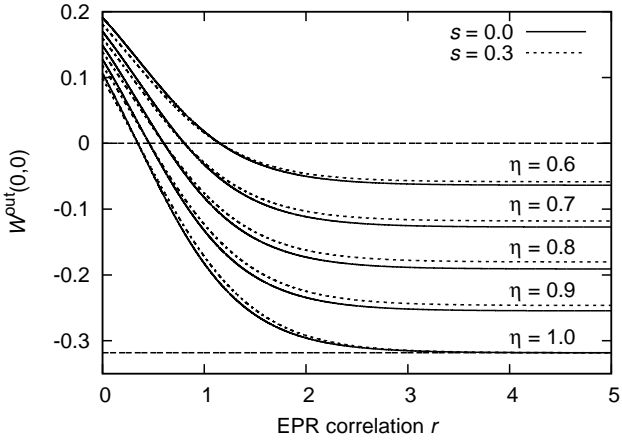}
  \caption{\label{fig_output_neg}Output Wigner function negativity $\Win(0,0)$ as a function of $r$ for different values of $\eta$ and $s$. For a given $\eta$, different $s$ curves cross the $W(0,0) = 0$ at the same $r(\eta)$.}
\end{figure}
In practical terms, this means that both operations do not commute and losses at the input stage have more effect on the quality of the overall process than losses at the output stage.
The output center negativity is now expressed as
\begin{equation}
  \Wout(0,0) = \frac{g_r(g_r-2\eta)}{\pi\left(g^2_r+4\eta(g_r-\eta)\text{sh}^2(s)\right)^{3/2}}, 
  \label{phasespace_neg_output}
\end{equation}
with $g_r = 1+2e^{-2r}$(plotted in Fig. \ref{fig_output_neg}). As expected for unity gain teleportation, the $\Wout(0,0) = 0$ threshold is still independent of the squeezing parameter $s$ and can be expressed as a function of the two parameters $\eta$ and $r$ alone by the simple relation (Fig. \ref{fig_neg_thre})
\begin{equation}
r = \ln \sqrt{2/(2\eta-1)} \quad \text{at threshold}.
\label{neg_threshold}
\end{equation}

\begin{figure}[!h]
  \includegraphics[height=0.48\columnwidth,angle=-90]{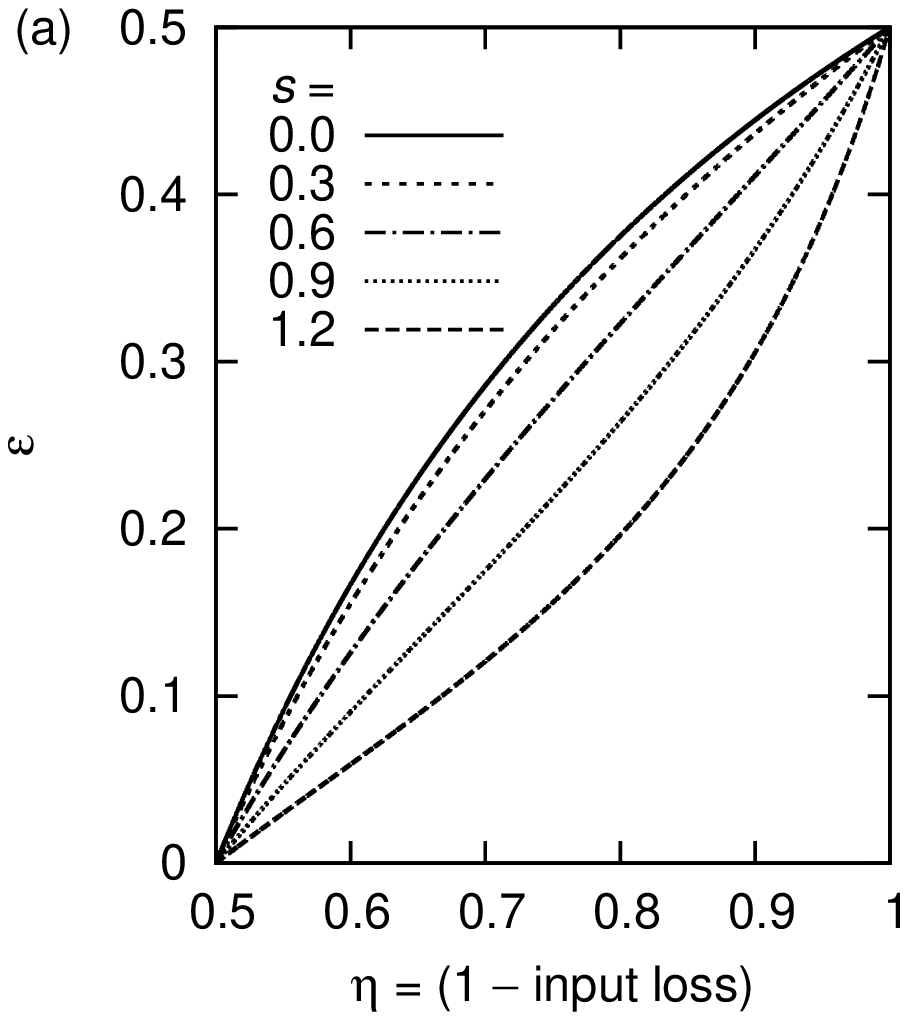}
  \includegraphics[height=0.48\columnwidth,angle=-90]{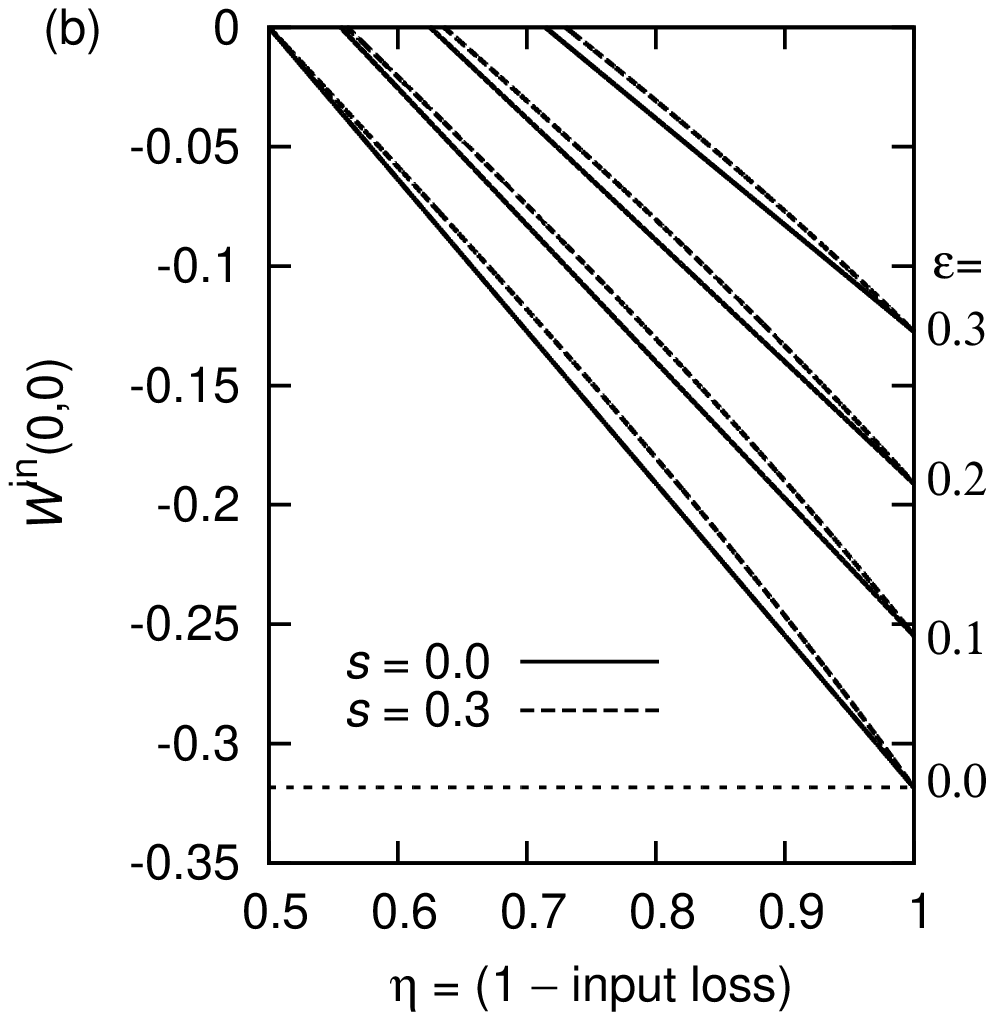}
  \caption{\label{fig_mix_input_neg2}(a) Input negativity threshold $\Win(0,0) = 0$ as a function of $\eta$, $s$, and $\epsilon$. (b) Input negativity $\Win(0,0)$ as a function of $\eta$ for different values of $s$ and $\epsilon$.}
\end{figure}

\begin{figure}[!h]
  \includegraphics[height=\columnwidth,angle=-90]{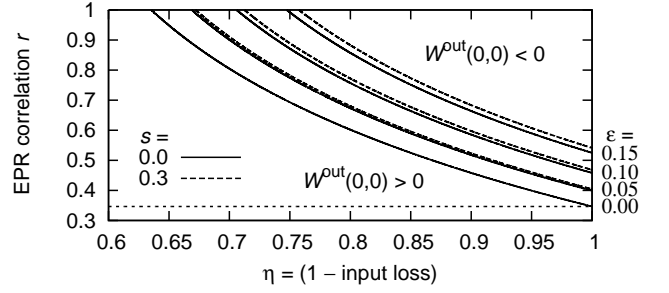}
  \caption{\label{fig_mix_output_neg}Influence of $\epsilon$ and $s$ on the negativity threshold $\Wout(0,0) = 0$ at the output of teleportation.}
\end{figure}

Until now, our model of reference state has assumed a pure state with an initial density matrix $\drho_\reflab$ of the form
\begin{equation}
\drho_\reflab = \anni \hat{S}_s \ketvac\bravac \hat{S}_{-s}\crea.
\end{equation}
However, to be more faithful to experimentally produced squeezed photon states, rather than an ideal photon resolving detector, we should consider the unideal projection properties of the Geiger silicon Avalanche PhotoDiode (APDs) experimentally used to produce photon subtracted states.
Essentially two mixing mechanisms are at work.
First, the on/off character of the APD makes it only able to detect the presence of photons without resolving the actual number of them.
This leads quite naturally to a Positive Operator Valued Measure (POVM) solution to model the APD measurement, as was done in \cite{Suzuki06}.
However, the effect is rather marginal if we restrict ourselves to small squeezing $s$ and small tapping fraction $R$ and we will actually neglect photon components higher than $n=1$ in the trigger channel.
Second, the laboratory APD is also characterized by a dark count rate, which will produce false heralding events and induce some statistical mixing of the target state.
When such a false event occurs, no projection happens on the signal mode as no photon is subtracted and the signal mode density matrix is just $\drho_\flslab = \hat{S}_s \ketvac\bravac \hat{S}_{-s}$ corresponding to the OPO output squeezed vacuum.
We introduce a parameter $\epsilon$ that reflects this statistical mixing and write the new reference density matrix $\drho_\reflab'$ as
\begin{equation}
\drho_\reflab' = (1-\epsilon) \drho_\reflab + \epsilon \drho_\flslab,
\label{mixed_ref_state}
\end{equation}
where $\epsilon$ is related to the modal purity parameter $\Xi$ introduced in \cite{Wenger04} by $1-\epsilon = \Xi$. 
With $\epsilon\neq 0$, the value of $\Wref(0,0)$ is not optimal anymore, but becomes (Fig. \ref{fig_mix_input_neg2}) 
\begin{eqnarray}
\Wref(0,0) \rightarrow 
& (1-\epsilon).\Wref(0,0) + \epsilon.W^\flslab(0,0) \nonumber\\
& = (2\epsilon-1)/\pi.
\end{eqnarray}
Correcting for the effect of Eq. \eqref{mixed_ref_state} in the input negativity \eqref{phasespace_neg_input}, output negativity \eqref{phasespace_neg_output}, and negativity threshold \eqref{neg_threshold} is just a matter of calculating how the Wigner function associated to $\drho_\flslab$, $W_\flslab(x,p) = G_{1/\sqrt{2}}(e^{s}x,e^{-s}p)$, evolves in the teleportation process.
This is possible since all the transformations used until now have been linear and therefore we can write
\begin{eqnarray}
W^\alpha(0,0)
 \rightarrow (1-\epsilon)W^\alpha(0,0) 
\qquad\qquad\qquad\nonumber\\
+ \frac{\epsilon}{\eta} (W_\flslab\circ G_{\lambda^\alpha})(0,0)\qquad,
\label{povm_transformation}
\end{eqnarray}
with $\alpha\in\{\inlab,\outlab\}$, $\lambda^\inlab = \lambda$, and $\lambda^\outlab = \lambda'$.
We find for the corrected input negativity the new expression
\begin{eqnarray}
\label{input_neg_with_mixing}
\Win(0,0) \rightarrow \, \Win(0,0) 
\qquad\qquad\qquad\qquad\nonumber\\ 
+ 2\epsilon\eta\frac{1+2(1-\eta)\text{sh}^2(s)}{\pi\left(1+4\eta(1-\eta)\text{sh}^2(s)\right)^{3/2}},
\end{eqnarray}
as well as
\begin{eqnarray}
\label{output_neg_with_mixing}
\Wout(0,0) \rightarrow \, \Wout(0,0) 
\qquad\qquad\qquad\qquad\nonumber\\ 
+ 2\epsilon\eta\frac{g_r+2(g_r-\eta)\text{sh}^2(s)}{\pi\left(g_r^2+4\eta(g_r-2\eta)\text{sh}^2(s)\right)^{3/2}},
\end{eqnarray}
for the corrected output negativity.
Input state negativity threshold $\Win(0,0) = 0$ now gives the following relation between $\eta$ and $\epsilon$
\begin{equation}
\epsilon = (2\eta-1)/2\eta\left(1+2(1-\eta)\text{sh}^2(s)\right).
\end{equation}
For nonzero $\epsilon$, the input threshold becomes dependent on the squeezing parameter $s$.
We expect the same dependence on the output negativity threshold corrected for $\epsilon$, which is now expressed with the following quadratic equation: 
\begin{equation}
0 = g_r^2 + 2\,b(\epsilon,s)\,\eta\,g_r-c(\epsilon,s)\,\eta^2,
\label{mix_neg_thre_quadratic}
\end{equation}
with $b(\epsilon,s) = \epsilon(1+2\text{sh}^2(s))-1$ and $c(\epsilon,s) = 4\epsilon\text{sh}^2(s)$.
Keeping the only physical solution in \eqref{mix_neg_thre_quadratic} the negativity teleportation threshold corrected for $\epsilon$ becomes (Fig. \ref{fig_mix_output_neg})
\begin{equation}
r = \text{ln} \left( \frac{2}{\eta\left( \sqrt{b^2+c}-b \right)-1}  \right)^{1/2}.  
\end{equation}

In summary, we have developed in this section a realistic yet simple model to account for the phase-space properties of photon subtracted squeezed vacuum states.
With that model we have considered the effect of teleportation on these states and are able to predict the success of negativity teleportation.
In the next section we will now show how to take into account the multi-mode aspect of these input states and the multi-mode aspect of broadband teleportation.


\section{Multimode teleportation}
\label{section_multi}

The APD triggered non-Gaussian state has been shown to have complex multimode properties \cite{Sasaki06,Molmer07}.
On one hand, the OPO output beam is a continuous wave with a specific squeezing spectrum, while, on the other hand, the APD triggers that herald a non-Gaussian state happen at precisely defined times.
Intuitively one expects that a given APD trigger induces non-Gaussian statistics in the immediate time vicinity of the trigger event to the extent of the OPO bandwidth. 
When the number of triggering events increases as $R$ and $s$ increase and more photons end up in the trigger channel, more complex time interference phenomena arise on the signal mode between neighboring photon subtracted wave packets\cite{Molmer07bis}.
Fortunately, in the limit of small $s$ and small $R$, a simple two modes picture allows one to efficiently describe the input state and capture most of its experimental properties \cite{Molmer06}.
It involves two effective wave-packet modes, $\hat{A}_s$ for the signal mode and $\hat{A}_t$ for the trigger mode, defined by 
\begin{equation}
\hat{A}_i = \int f_i(\omega) \anni_{\omega} d\omega,
\label{wavepacket_mode}
\end{equation}
with $i\in\{s,t\}$. 
Preserving the commutators $[\hat{A}_i,\hat{A}^\dagger_i] = 1$ requires $\int |f_i(\omega)|^2 d\omega = 1$.
While the exact form of $f_t(\omega)$ is not really relevant, since the APD detection time is typically much shorter than any other time scale in these continuous wave experiments, $f_s(\omega)$ will describe the temporal characteristic of the heralded non-Gaussian state.
This function will be defined by the OPO bandwidth, as well as the possible filtering cavities used on the trigger channel and numerical optimization has shown that the optimal form (small $s$ and $R$, wide filtering cavities) can be taken as 
\begin{equation}
f(\omega) = \gamma/\pi[\gamma^2+(\omega_0-\omega)^2],
\end{equation}
with $\gamma$ the OPO decay rate including intra-cavity losses and $\omega_0$ the light beam carrier frequency\cite{Molmer07bis}.
In short, an APD heralded state behaves as a traveling wave packet of light with non-Gaussian characteristics.
Using the definition of $\hat{A}_s$ in Eq. \eqref{wavepacket_mode}, we return to a pure single-mode model for the input state that we write as
\begin{equation}
\ket{\psi} = e^{-s({\hat{A}^{\dag 2}}-\hat{A}^2)/2}\hat{A}^\dag\ketvac.
\label{wavepacket_mode_input}
\end{equation}
This expression is still an approximation of reality in the sense that a true multimode description as done in \cite{Sasaki06} would require an extension of a broadband squeezing operator $\hat{S}_B$ on a basis of adequately chosen orthogonal functions $\{f_n(\omega)\}$.
In this case, $\hat{S}_B$ is expressed as
\begin{equation}
\hat{S}_B = \exp \left[ - \int \frac{d\Omega}{2\pi}\frac{\zeta(\Omega)}{2} 
\left( \hat{A}^\dagger_{\Omega}\hat{A}^\dagger_{-\Omega}-\hat{A}_{\Omega}\hat{A}_{-\Omega} \right) \right],
\end{equation}
with $\hat{A}_\Omega = \anni(\omega_0+\Omega)$.
In practice, once a signal mode $f_s = f_0$ is chosen, the other modes $n \neq 0$ are traced out, which leads to mixing of the density matrix to the extent of the multimode entanglement present in $\hat{S}_B$ between all modes $n$:
\begin{equation}
\hat{\rho}_B = \text{tr}_{n\neq0}
\left( \hat{A}_0\hat{S}_B |0\rangle\langle0|\hat{S}_B^\dagger \hat{A}_0^\dagger \right).
\end{equation}
This multimode entanglement is quantitatively tracked by the function $\zeta(\Omega)$.
A special case happens when $\zeta$ is a constant at every frequency.
Then the operator $\hat{S}_B$ can be exactly factorized on the orthogonal basis $\{f_n(\omega)\}$ and the expression \eqref{wavepacket_mode_input} becomes exact.
In the general case of a non constant function $\zeta$, the expression \eqref{wavepacket_mode_input} is nevertheless useful as it only neglects a small amount of entanglement between the different orthogonal modes $n$ if $s$ is small. 
As a result, we will use expressions \eqref{wavepacket_mode} and \eqref{wavepacket_mode_input} for our following analysis of multimode teleportation.

To investigate how this multimode aspect translates quantitatively, we express in the Heisenberg picture the relation between input $(\xo_{\text{in}},\po_{\text{in}})$ and output $(\xo_{\text{out}},\po_{\text{out}})$ quadrature operators of teleportation as
\begin{eqnarray}
\xo_{\text{out}} = g_x\xo_{\text{in}} 
- \frac{1+g_x}{\sqrt{2}} e^{-r} \hat{v}_x
+ \frac{1-g_x}{\sqrt{2}} e^{+r} \hat{w}_x,
\label{tele_mono_heisenberg_anygain_bis}
\\
\po_{\text{out}} = g_p\po_{\text{in}} 
+ \frac{1+g_p}{\sqrt{2}}  e^{-r} \hat{w}_p
- \frac{1-g_p}{\sqrt{2}}  e^{+r} \hat{v}_p,
\label{tele_mono_heisenberg_anygain}
\end{eqnarray}
with $(\hat{v}_x,\hat{v}_p)$ and $(\hat{w}_x,\hat{w}_p)$ two auxiliary modes in the vacuum state\cite{vanLoock05}. 
We first consider the unity gain case $g_x = g_p = 1$, where the input/output relations \eqref{tele_mono_heisenberg_anygain_bis} and \eqref{tele_mono_heisenberg_anygain} simplify to
\begin{equation}
\xo_{\text{out}} = \xo_{\text{in}} - \sqrt{2} e^{-r} \hat{v}_x, 
\po_{\text{out}} = \po_{\text{in}} + \sqrt{2} e^{-r} \hat{w}_p.
\label{tele_mono_heisenberg}
\end{equation}
We notice that the input modes $\xo_{\text{in}}$, $\po_{\text{in}}$ and output modes $\xo_{\text{out}}$, $\po_{\text{out}}$ can actually represent any frequency mode $\omega$ and we
define in the same way as in Eq. \eqref{wavepacket_mode} two new modes, the input $\hat{A}_\inlab$ and output $\hat{A}_\outlab$ wave-packet modes by
\begin{equation}
\hat{A}_\inlab = \int f_s(\omega) \anni_\inlab({\omega}) d\omega, 
\hat{A}_\outlab = \int f_s(\omega) \anni_\outlab({\omega}) d\omega.
\end{equation}
as well as the input and output wave-packet quadratures $(\hat{X}_{\text{in}},\hat{P}_{\text{in}})$ and $(\hat{X}_{\text{out}},\hat{P}_{\text{out}})$ relevant for wave-packet teleportation.
We can directly rewrite the teleportation input/output relationship \eqref{tele_mono_heisenberg} in the form
\begin{eqnarray}
\label{tele_multi_heisenberg_A}
\hat{X}_{\text{out}} 
= \hat{X}_{\text{in}} - \sqrt{2} \int f_s(\omega) e^{-r(\omega)} \hat{v}_x\, d\omega,\\
\hat{P}_{\text{out}} 
= \hat{P}_{\text{in}} + \sqrt{2} \int f_s(\omega) e^{-r(\omega)} \hat{w}_p\, d\omega,
\end{eqnarray}
where we have introduced $r(\omega)$ as the spectrum of EPR correlations resolved in frequency.
From the physical properties of the OPO cavities used for EPR squeezing generation it is possible to deduce the expression of $r(\omega)$ from the squeezing spectrum $S_-(\omega) = \langle \Delta^2 \hat{x}^\dagger_\text{sqd}(\omega) \Delta^2 \hat{x}_\text{sqd}(\omega) \rangle / \langle \Delta^2 \hat{x}_\text{vac} \rangle $.
See, for example, Ref. \cite{Loock00} for details.
We now define an effective broadband EPR parameter $r_{\text{eff}}$ by
\begin{equation}
\label{effective_correlation}
 e^{-r_{\text{eff}}} 
= \int f_s(\omega)  e^{-r(\omega)}  d\omega 
= \int f_s(\omega) S_-(\omega) d\omega,
\end{equation}
and therefore Eqs. \eqref{tele_multi_heisenberg_A} simplify themselves to 
\begin{equation}
\label{tele_multi_heisenberg_B}
\hat{X}_{\text{out}} = \hat{X}_{\text{in}} - \sqrt{2} e^{-r_{\text{eff}}}\, \hat{v}_x, 
\hat{P}_{\text{out}} = \hat{P}_{\text{in}} + \sqrt{2} e^{-r_{\text{eff}}}\, \hat{w}_p.
\end{equation}
Since the auxiliary modes $\hat{v}_x$ and $\hat{w}_p$ are effectively traced out on the vacuum state at all frequencies, it is possible to take them outside of the frequency domain integrals to obtain the formulation \eqref{tele_multi_heisenberg_B}.
We notice that Eqs. \eqref{tele_multi_heisenberg_B} are written in the same way as Eqs. \eqref{tele_mono_heisenberg}. 
Thanks to the linearity of transformation \eqref{tele_mono_heisenberg} and the linear model of input state \eqref{wavepacket_mode} in the Heisenberg picture, multimode teleportation is equivalent to familiar single-mode teleportation, where an effective broadband EPR parameter $r_{\text{eff}}$ has been defined to take into account the finite bandwidth of entanglement.
In short, all the previous formulas of Sec. \ref{section_input} for unity-gain teleportation are readily usable with the simple change $r\rightarrow r_{\text{eff}}$.

The case of nonunity gain multimode teleportation is much more complex and we conclude this section with a brief overview of the non-unity gain case.
First we introduce the transfer functions $g_x(\omega)$ and $g_p(\omega)$, which represent the effects in frequency space of the classical channel.
$g_x$ and $g_p$ are in general complex-valued functions verifying the Kramers-Kronig relations.
We are now facing the problem that the output quadratures operators of teleportation will not be Hermitian operators anymore, in general. 
By taking an approach similar to Eq. \eqref{tele_multi_heisenberg_A}, we obtain for the position quadrature
\begin{eqnarray} 
\label{nonunity_multimode_heisenberg_A}
\hat{X}_\text{out}^\text{nonunit} 
 = \int f_s(\omega) g_x(\omega) \hat{x}_{\text{in}}(\omega) d\omega 
\qquad\qquad\qquad\qquad \nonumber\\
\quad - \frac{1}{\sqrt{2}} \left( e^{-r_{\text{eff}}} 
+ \int f_s(\omega) g_x(\omega) e^{-r(\omega)} d\omega\right) \hat{v}_x       
\qquad\, \nonumber\\
\quad + \frac{1}{\sqrt{2}} \left( e^{+r_{\text{eff}}} 
- \int f_s(\omega) g_x(\omega) e^{+r(\omega)} d\omega\right) \hat{w}_x. \qquad
\end{eqnarray}
where $\hat{X}_\text{out}^\text{nonunit} $ is the output teleported mode in the nonunity gain regime. 
To clean this expression we define the two complex numbers $g^\pm_x$:
\begin{equation}
\label{nonunity_multimode_heisenberg_cleaning}
g^\pm_x = \int f_s(\omega)g_x(\omega)e^{ \pm\left( r(\omega)-r_{\text{eff}} \right) } d\omega,
\end{equation}
so that Eq. \eqref{nonunity_multimode_heisenberg_A} simplifies to
\begin{eqnarray}
\label{nonunity_multimode_heisenberg_B}
\hat{X}_{\text{out}}^\text{nonunit} 
= \int f_s(\omega) g_x(\omega) \hat{x}_{\text{in}}(\omega) d\omega 
\qquad\qquad\qquad\nonumber\\
\qquad - \frac{1 + g^-_x}{\sqrt{2}} e^{-r_{\text{eff}}} \hat{v}_x
       + \frac{1 - g^+_x}{\sqrt{2}} e^{+r_{\text{eff}}} \hat{w}_x, \qquad
\end{eqnarray}
with a similar expression for $\hat{P}_{\text{out}}^\text{nonunit}$.
By further separating $\hat{X}_\text{out}^\text{nonunit}$ and $\hat{P}_{\text{out}}^\text{nonunit}$ in real and imaginary parts as done in \cite{Loock00}, it is possible to obtain from this model observable results.
Equations \eqref{nonunity_multimode_heisenberg_B} and \eqref{nonunity_multimode_heisenberg_cleaning} show that the output modes will get contaminated by antisqueezing when $|g_x(\omega)|$ and $|g_p(\omega)|$ are different from 1.
Furthermore, the expression $\int f_s(\omega) g_x(\omega) \hat{x}_{\text{in}}(\omega) d\omega$ hints that the wave-packet shape from input to output will get modified by the teleportation process.
An interesting and practical situation is the case of pure linear delay  $g_x(\omega) = g_p(\omega) = \exp[-i\omega\,\Delta t]$.
If such a phase factor is added by the classical channel to the output modes, its full effect can be absorbed in $\hat{x}_\text{in}$ and auxiliary modes $\hat{v}_x$, $\hat{w}_x$ by using their Fourier transforms,
\begin{equation}
\hat{x}_\text{in}(\omega) 
\rightarrow \hat{x}_\text{in}(\omega) e^{-i\omega\,\Delta t} \nonumber\\
= \frac{1}{\sqrt{2\pi}} \int dt \hat{x}_\text{in}(t) e^{i\omega(t-\Delta t)},
\end{equation}
so that $\hat{X}_{\text{out}}^\text{nonunit}$ is related to $\hat{X}_{\text{out}}$ by a simple time translation
\begin{equation}
\hat{X}_{\text{out}}^\text{nonunit}(t) = \hat{X}_{\text{out}}(t+\Delta t).
\end{equation}
As a matter of fact, this is exactly how the experimental teleportation setup used in \cite{Lee11} behaves, where an optical delay line is used to match the phase answer of the classical channel and cancel this $\Delta t$ phase factor.

In this section, we have developed an efficient model of unity-gain multimode teleportation with the added benefit of being able to use all the results of the previous section.
In the next section we will further use this model to investigate the effects of classical sources of noise in the classical channel and their effect on the teleportation process.


\section{Noise model}
\label{section_noise}

In this section, we try to better understand the effect of classical sources of noise on the teleportation process.
This is an important point to consider as, compared to the single sideband regime, it is much harder to experimentally insulate from external noise a broad range of frequencies at the same time. 
To throw light on that issue, we first look for a master equation describing the effect of teleportation on the density matrix $\drho$. 
For that, we start by fully detailing Eq. \eqref{phasespace_tele} with input $W$ and output $W'$:
\begin{equation}
W'_{(x,p)} = \frac{1}{2\pi\sigma^2} \int\int dx'dp' W_{(x',p')} e^{-\frac{(x-x')^2+(p-p')^2}{2\sigma^2}}.
\end{equation}
We now assume that $\sigma \rightarrow \sigma(t)$ has a time dependence.
We express the first derivative of $W'$ with respect to time $t$, 
\begin{eqnarray}
\frac{d}{dt} W'_{(x,p)} 
= \frac{2}{\sigma(t) } \frac{d}{dt} \sigma(t) \left( -W'_{(x,p)}\right.
\qquad\qquad\qquad\qquad\nonumber\\
+\frac{1}{2\sigma^2} \int\int \frac{dx'dp'}{2\pi\sigma^2} W_{(x',p')} 
\qquad\qquad\qquad\qquad\,\nonumber\\
\left.\times (\,(x-x')^2+(p-p')^2\,) e^{-\frac{(x-x')^2+(p-p')^2}{2\sigma^2}} \right),
\quad
\end{eqnarray}
and the second derivative of $W'$ with respect to position $x$
\begin{eqnarray}
\derv{x}^2 W'_{(x,p)} = -\frac{1}{\sigma^2} W'_{(x,p)} + \frac{1}{\sigma^4} \int\int \frac{dx'dp'}{2\pi\sigma^2}\nonumber\\
\quad \times (x-x')^2 W_{(x',p')} e^{-\frac{(x-x')^2+(p-p')^2}{2\sigma^2}}.
\end{eqnarray}
We immediately find the following differential equation for $W'_{(x,p;t)}$,
\begin{equation}
\frac{d}{dt} W'_{(x,p;t)} = \frac{1}{2}\left(\derv{t}\sigma^2(t)\right)\Delta W'_{(x,p;t)},
\label{diff_eq_2}
\end{equation}
where $\Delta = \derv{x}^2 + \derv{p}^2$. 
We choose $\sigma$ to be $\sigma(t) = \sqrt{2\kappa't}$ with $\kappa'$ a constant decay rate, so that Eq. \eqref{diff_eq_2} simplifies itself to a pure diffusion equation,
\begin{equation}
\frac{d}{dt} W'_{(x,p;t)} = \kappa' \Delta W'_{(x,p;t)}.
\label{diff_eq_3}
\end{equation}
Then, by using correspondence rules between the phase space formalism and the density matrix formalism\cite{Gardiner04}, we find from Eq. \eqref{diff_eq_3} the following master equation for $\drho$:
\begin{eqnarray}
\frac{d}{dt}\drho = \kappa' \left( 2\crea\drho\anni + 2\anni\drho\crea \right.
\qquad\qquad\nonumber\\ 
\qquad\qquad\left.- \crea\anni\drho - \anni\crea\drho - \drho\crea\anni - \drho\anni\crea \right),
\label{master_2}
\end{eqnarray}
which can be equivalently written as
\begin{equation}
\frac{d}{dt}\drho = L[\drho] , \quad L[\drho] = \kappa'[\crea,[\drho,\anni]] + \kappa'[\anni,[\drho,\crea]].
\end{equation}
The master equation \eqref{master_2} is the well-known damping process for the harmonic oscillator.
Equations \eqref{tele_mono_heisenberg} look therefore similar to quantum Langevin equations, where the terms $\sqrt{2} e^{-r} \hat{v}_x$ and $\sqrt{2} e^{-r} \hat{v}_p$ are nothing other than thermalization terms.
We would intuitively add the effect of any classical source of noise directly in Eqs. \eqref{tele_mono_heisenberg} by writing
\begin{eqnarray}
\label{heisenberg_teleportation_with_noise}
\xo_{\text{out}} = \xo_{\text{in}} - \sqrt{2} e^{-r} \hat{v}_x - \sqrt{2}\,\N_x\,\hat{y},
\nonumber\\
\po_{\text{out}} = \po_{\text{in}} + \sqrt{2} e^{-r} \hat{w}_p + \sqrt{2}\,\N_p\,\hat{z},
\end{eqnarray}
where we have introduced two new auxiliary vacuum modes $\hat{y}$ and $\hat{z}$ and where we first are considering the single-mode case.
$\N_x$ and $\N_p$ describe the amplitude of noise normalized to vacuum and added at the output of teleportation on top of finite squeezing.
This noise can arise independently for both quadratures from imperfect electronics in the classical channel, for example.
The most natural case is for noise to be uncorrelated with quadrature angle and we can assume $\N_x = \N_p = \N$ to be the average noise amplitude.
We remember that all auxiliary modes appearing in the Heisenberg picture teleportation equations are traced out on the vacuum state and are uncorrelated.
Therefore it would be natural to redefine a correlation parameter $r'$ modified by the amount of noise with the simple relation
\begin{equation}
 e^{-r} \rightarrow  e^{-r'} = e^{-r} + \N,
\end{equation}
so that Eq. \eqref{heisenberg_teleportation_with_noise} would be written as Eq. \eqref{tele_mono_heisenberg}. 
However, this approach is wrong and the correlation parameter $r'$ cannot be redefined in amplitude but should be redefined in power by writing
\begin{equation}
 e^{-2r} \rightarrow  e^{-2r'} = e^{-2r} + \N^2.
\end{equation}
It is possible to justify this expression rigorously by establishing the link between Heisenberg picture equations \eqref{tele_mono_heisenberg} and the original phase space formulation of Eq. \eqref{phasespace_tele}.
For that purpose, we introduce the characteristic function $\Chi(\alpha)$ related to the density matrix $\drho$ by the Weyl expansion formula
\begin{equation}
\Chi(\alpha) = \text{tr}(\drho\hat{D}_{\alpha}) = \langle \hat{D}_{\alpha} \rangle \,,\quad \drho = \int d\alpha \Chi(\alpha) \hat{D}_{-\alpha},
\label{char_function}
\end{equation}
where $\hat{D}_\alpha$ is the displacement operator $\text{exp}[\alpha\crea-\alpha^*\anni]$.
If we write $\alpha = (u+iv)/\sqrt{2}$, then $\Chi$ and $W$ are related by the following Fourier transform:
\begin{equation}
W(x,p) = \frac{1}{4\pi^2}\int\int dudv \Chi(u,v) e^{ivx-iup}.
\label{char_to_wigner}
\end{equation}
Now if we consider the unitary transformation
\begin{equation}
\xo \rightarrow \xo' = \xo - \sqrt{2}\gamma \hat{v}_x, 
\end{equation}
with an auxiliary mode $(\hat{v}_x,\hat{v}_p)$ having commutators $[\xo,\hat{v}_x] = [\po,\hat{v}_x] = 0$ and $[\hat{v}_x,\hat{v}_p]=i$, then the displacement operator $\hat{D}_\alpha$ is changed to
\begin{equation}
\hat{D}_\alpha \rightarrow \hat{D}_\alpha' = \hat{D}_\alpha \otimes \hat{D}^{\hat{v}}_{\alpha'} 
=  \hat{D}_\alpha \otimes e^{+i\sqrt{2}u\gamma\hat{v}_p},
\end{equation}
with $\hat{D}^v$ a displacement operator acting on mode $\hat{v}$ and $\alpha' = (-\sqrt{2}\gamma \times u+i\times 0)/\sqrt{2}$.
To express the new characteristic function $\Chi'$, we have to evaluate the trace of $\hat{D}^v$ taken on the vacuum for mode $\hat{v}$:
\begin{equation}
\text{tr}\left(\ketvac\bravac \hat{D}^v_{\alpha'}\right) 
= \bravac \hat{D}^v_{\alpha'} \ketvac 
= e^{-|\alpha'|^2/2}.
\end{equation}
This lead to the expression of $\Chi'(u,v)$,
\begin{equation}
\Chi'(u,v) = \Chi(u,v) e^{-\gamma^2u^2/2},
\label{char_gaussian}
\end{equation}
which immediately translates to a Gaussian convolution, such as Eq. \eqref{phasespace_tele}, for the Wigner function $W$ having the Fourier relationship \eqref{char_to_wigner} between $\Chi$ and $W$.
In this case, we obtain the semiconvolution
\begin{equation}
W'(x,p) = \frac{1}{\sqrt{2\pi}\gamma} \int dx' W(x',p) e^{-(x-x')/2\gamma^2}.
\end{equation}
By also adding the transformation $\po \rightarrow \po' = \po + \sqrt{2} \gamma \hat{w}_p$, we would finally obtain Eq. \eqref{phasespace_tele} provided we define $\gamma$ as equal to $\exp[-r]$. 
If we now also consider the added noise term $\sqrt{2}\N\hat{y}_n$ in Eqs. \eqref{heisenberg_teleportation_with_noise}, $\Chi'(u,v)$ would be written 
\begin{equation}
\Chi'(u,v) = \Chi(u,v) e^{-(\gamma^2 + \N^2 )u^2/2},
\end{equation}
which justifies to redefine the correlation parameter $r$ in power and not in amplitude as
\begin{equation}
\label{correlation_with_noise}
r \rightarrow r' = r - \ln \sqrt{1+\N^2 e^{2r}}.
\end{equation}
\begin{figure}[!h]
  \includegraphics[height=\columnwidth,angle=-90]{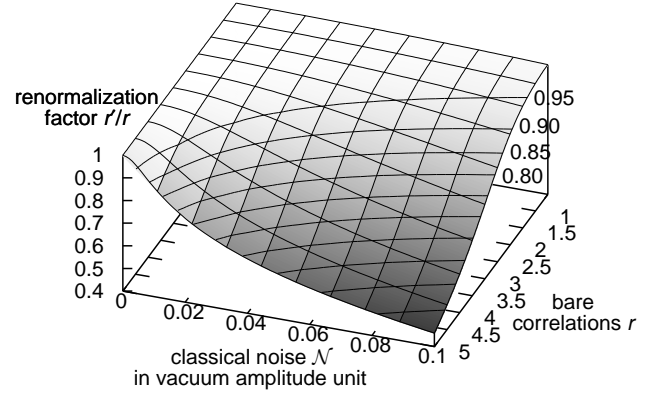}
  \caption{\label{classical_noise_fig_neg}Influence of the noise amplitude $\N$ on the ratio $r'$ and $r$.}
\end{figure}
We see that if the amount of noise $\N$ is high, it is possible that $r'$ becomes negative (see Fig. \ref{correlation_with_noise}), which simply means that after factoring in the effect of $\N$, quantum teleportation would perform worse than classical teleportation with $r=0$ and $\N=0$.
More interesting is the case of broadband noise when $\N \rightarrow \N(\omega)$ contaminates the whole frequency range relevant for teleportation.
In the same way that we had deduced Eqs. \eqref{tele_multi_heisenberg_B} from Eqs. \eqref{tele_mono_heisenberg} using the wave-packet operator \eqref{wavepacket_mode}, we define an effective noise level $\N_\text{eff} = \int f_s(\omega) \N(\omega) d\omega$ and write
\begin{eqnarray}
\hat{X}_{\text{out}} 
= \hat{X}_{\text{in}} 
- \sqrt{2} e^{-r_{\text{eff}}} \hat{v}_x
- \sqrt{2}\, \N_\text{eff}\, \hat{y}_n, \nonumber\\
\hat{P}_{\text{out}} 
= \hat{P}_{\text{in}} 
+ \sqrt{2} e^{-r_{\text{eff}}} \hat{w}_p
+ \sqrt{2}\, \N_\text{eff}\, \hat{z}_n,
\end{eqnarray}
so that finally it is possible to take into account the effect of $\N(\omega)$ by redefining $r_\text{eff}$ as in Eq. \eqref{correlation_with_noise}:
\begin{equation}
\label{correlation_with_noise}
r_\text{eff} \rightarrow r'_\text{eff} = r_\text{eff} - \ln \sqrt{1+\N_\text{eff}^2 e^{2r_\text{eff}}}
\end{equation}

In this section, we have shown how to estimate the effect of external classical noise sources on the output teleported modes.
This simple model only works for Gaussian entanglement and for sources of uncorrelated Gaussian noise.
By Gaussian noise, we mean that the underlying quantum state used to trace out auxiliary quantum modes $\hat{y}$ and $\hat{z}$ is Gaussian.
In this case, external classical noise becomes essentially indistinguishable from noise added by the teleportation itself due to finite squeezing.
Furthermore our model is able to take into account any spectrum of noise $\N(\omega)$ by using the wave-packet mode function $f_s$ to estimate an effective level of noise $\N_\text{eff}$ added to the teleportation.


\section{Back-testing and conclusion}
\label{section_conclu}

The first step to test the validity of our results is to check that the model of Sec. \ref{section_input} we used for photon subtracted squeezed vacuum states works well with the experimental input states used in \cite{Lee11}.
For that we need to estimate three parameters: the squeezing parameter $s$, the loss parameter $1-\eta$, and the APD dark noise parameter $\epsilon$.
In \cite{Lee11}, a direct measure of the APD dark noise and event counts gives for $\epsilon$ a value of 0.013.
With the help of quantum tomography of the input squeezed vacuum state $\hat{S}_{s}\ketvac$ the squeezing parameter $s$ is estimated to be 0.28.
This tomography is done using the wave-packet function $f_s$ as a filter of the measured homodyne currents and without conditioning on the APD triggers.
This means that $s$ is actually an effective squeezing parameter in the sense of Eq. \eqref{effective_correlation}, taking into account the bandwidth of the OPO used to generate the state $\hat{S}_{s}\ketvac$.
Finally, to estimate the value of $\eta$, we use the equation \eqref{input_neg_with_mixing} with the value of $\Win(0,0)$ obtained from a quantum tomography of the input state and obtain $\eta = 0.80$.
This value is slightly different from the one estimated in \cite{Lee11} due to the non-zero value of $\epsilon$.
With these three parameters known we can numerically simulate $\Win$ using the results of Sec. \ref{section_input} and compare it to the reconstructed Wigner function with the overlap formula,
\begin{equation}
\label{wigner_overlap}
O(W_a,W_b) = 2\pi \int\int dxdp W_a(x,p) W_b(x,p).
\end{equation}
However, this formula does not work so straightforwardly in our case: 
if $W_a$ and $W_b$ are mixed states and even though $W_a = W_b$, the overlap given by formula \eqref{wigner_overlap} will not be 1 but rather the purity of $W_a$.
We therefore use a modified version of the above formula with a renormalization factor taking into account the purity of both quantum states:
\begin{equation}
\label{wigner_overlap_normed}
O'(W_a,W_b) = \frac{O(W_a,W_b)}{\left(O(W_a,W_a)O(W_b,W_b)\right)^{1/2}}.
\end{equation}
With this modified overlap formula \eqref{wigner_overlap_normed} we calculate an overlap of 0.987 between our model and the reconstructed state.
The $L2$ Euclidian distance $d(W_a,W_b)$ defined by 
\begin{equation}
d(W_a,W_b)
= \left(\int\int dxdp \,\left|W_a(x,p)-W_b(x,p)\right|^2\right)^{1/2},\quad
\end{equation}
between the two states is found to be 0.05.
Finally one could also choose to maximize $O'$ rather than fitting the value of $\Win(0,0)$ to estimate $\eta$.
However, because this approach can lead to a value of $\Win(0,0)$ significantly different from the experimentally measured value, we chose to directly fit $\Win(0,0)$ instead.

The second step is to estimate the broadband EPR parameter $r_{\text{eff}}$, again choosing one of two possible methods.
A first method would consist of directly measuring spectra of the EPR correlations between Alice and Bob and then using the mode function $f_s$ to obtain an estimation of $r_{\text{eff}}$.
While this method automatically takes into account homodyne finite efficiency and phase errors, it does not probe any imperfections of the classical channel.
A second method would consist of estimating $r_{\text{eff}}$ with a measure of vacuum teleportation fidelity, with the added benefit of taking into account the whole process of teleportation.
In the simple case of the vacuum state $\ketvac$ as an input state, the teleportation fidelity and the EPR correlations parameter $r$ are directly related by the relation
\begin{equation}
F_\text{tele} = 1 / (1 + e^{-r}).
\end{equation}
To measure the fidelity $F_\text{tele}$ we first make a tomographic reconstruction of the teleported vacuum state.
To specifically estimate $r_{\text{eff}}$ we consider the wave-packet vacuum state $\ketvac_{f_s}$ defined by
\begin{equation}
\ketvac_{f_s} = \int d\omega f_s(\omega) \ketvac_\omega.
\end{equation}
As before this is simply done by using the wave-packet mode function $f_s$ as a filtering function in quantum tomography.
With this second method, $r_{\text{eff}}$ is estimated to be 0.795 in \cite{Lee11}.

Finally, the third and final step consists of checking our prediction of $\Wout$ and particularly $\Wout(0,0)$ using all the known parameters.
This is  done by evaluating Eq. \eqref{output_neg_with_mixing} and we obtain the value $\Wout(0,0) = -0.0243$, in agreement with the measured output negativity of $-0.022$ in \cite{Lee11}.
As before since $\epsilon \neq 0$ the value of $\Wout(0,0)$ estimated here is slightly different from the estimation found in \cite{Lee11}.
We also check the overlap $O'$ between the measured state and the predicted state and find a value of 0.988. 
The $L2$ distance between the two states is 0.04.

Overall, the results of Sec. \ref{section_input} are numerically in good agreement with the experimental results of \cite{Lee11}.
Our model uses a set of only three parameters and even though more complex models for photon subtracted squeezed vacuum states exist, we found it was not necessary to use them.
This confirms the validity of our initial assumption to only consider the small $R$, small $s$ regime.
Our approach to multimode teleportation in Sec. \ref{section_multi} has the double benefit of intuitively picturing broadband operations in term of wave packets, while at the same time allowing us to use the usual results of single-mode teleportation with simple renormalization prescriptions.

In conclusion, we have developed an efficient yet simple framework to model the properties of multimode continuous variable teleportation.
Although this work is tied to a specific class of non-Gaussian states, it is natural to ask if a similar approach can handle more general non-Gaussian states.
As it is known that any Wigner function can be approximated by successive displacements and photon subtractions, it is in principle possible to describe any non-Gaussian states in a systematic way that would be compatible with the Gaussian convolutions needed in Sec. \ref{section_input}.
A second and harder issue would be to identify criteria more robust than negativity to decide on the success of teleportation and Gaussian operations in general in the context of non-Gaussian non-classical input states.


\begin{acknowledgments}
This work was partly supported by the Strategic Information and Communications R\&D Promotion (SCOPE) program of the Ministry of Internal Affairs and Communications of Japan, Project for Developing Innovation Systems, Grants-in-Aid for Scientific Research, Global Center of Excellence, Advanced Photon Science Alliance, and Funding Program for World-Leading Innovative R\&D on Science and Technology (FIRST) commissioned by the Ministry of Education, Culture, Sports, Science and Technology of Japan, and ASCR-JSPS, the Academy of Sciences of the Czech Republic and the Japanese Society for the Promotion of Science.
\end{acknowledgments}


\bibliographystyle{aipproc}

\begin{thebibliography}{99}


\bibitem[Bennet and al(1993)]{Bennet93}
{C. H. Bennett, G. Brassard, C. Cr\'epeau, R. Jozsa, A. Peres, and W. K. Wootters, Phys. Rev. Lett \textbf{70}, 1895 (1993).}

\bibitem[Vaidman(1994)]{Vaidman94}
{L. Vaidman, Phys. Rev. A \textbf{49}, 1473 (1994).}

\bibitem[Bouwmeesterr and al(1997)]{Zeilinger97}
{D. Bouwmeester, J.-W. Pan, K. Mattle, M. Eibl, H. Weinfurter, and A. Zeilinger
 Nature (London), \textbf{390}, 575 (1997).}

\bibitem[Furusawa and al(1998)]{Furusawa98}
{A. Furusawa, J. L. S\o rensen, S. L. Braunstein, C. A. Fuchs, H. J. Kimble, and E. S. Polzik, Science \textbf{282}, 706 (1998).}

\bibitem[Yukawa and al(2008)]{Yukawa08}
{M. Yukawa, H. Benichi, and A. Furusawa, Phys. Rev. A \textbf{77}, 022314 (2008).}

\bibitem[Yoshikawa(2007)]{Yoshikawa07}
{J. Yoshikawa, T. Hayashi, T. Akiyama, N. Takei, A. Huck, U. L. Andersen, and A. Furusawa, Phys. Rev. A. \textbf{76}, 060301(R) (2007).}

\bibitem[Llyod and Braunstein(1999)]{Braunstein99}
{S. Lloyd and S. L. Braunstein, Phys. Rev. Lett \textbf{82}, 1784 (1999).}

\bibitem[Ourjoumtsev(2006)]{Ourjoumtsev06}
{A. Ourjoumtsev, R. Tualle-Brouri, J. Laurat, and P. Grangier, Science \textbf{312}, 83 (2006).}

\bibitem[Jonas and al(2006)]{Jonas06}
{J. S. Neergaard-Nielsen, B. M. Nielsen, C. Hettich, K. M\o lmer, and E. S. Polzik, Phys. Rev. Lett. \textbf{97}, 083604 (2006).}

\bibitem[Wakui and al(2007)]{Wakui07}
{K. Wakui, H. Takahashi, A. Furusawa, and M. Sasaki, Opt. Express \textbf{15}, 3568 (2007).}

\bibitem[Lee and al(2010)]{Lee11}
{N. Lee, H. Benichi, Y. Takeno, S. Takeda, J. Webb, E. Huntington, and A. Furusawa, Science \textbf{332}, 330 (2011).}

\bibitem[Dakna and al(2006)]{Dakna97}
{M. Dakna, T. Anhut, T. Opatrn\'y, L. Kn\"oll, and D.-G. Welsch, Phys. Rev. A \textbf{55}, 3184 (1997).}

\bibitem[Ban(2004)]{Ban04}
{M. Ban, Phys. Rev. A \textbf{69}, 054304 (2004).}

\bibitem[Mista and al(2010)]{Mista10}
{L. Mista Jr., R. Filip, and A. Furusawa, Phys. Rev. A \textbf{82}, 012322 (2010).}

\bibitem[Noh(2009)]{Noh09}
{C. Noh, A. Chia, H. Nha, M.J. Collett, and H.J. Carmichael, Phys. Rev. Lett \textbf{102}, 230501 (2009).}

\bibitem[van Loock(2000)]{Loock00}
{P. van Loock, S. L. Braunstein, and H. J. Kimble, Phys. Rev. A \textbf{62}, 022309 (2000).}

\bibitem[Braunstein and Kimble(1998)]{KimBrauns98}
{S. L. Braunstein and H. J. Kimble, Phys. Rev. Lett \textbf{80}, 869 (1998).}

\bibitem[Grosshans and Grangier(2001)]{Grosshans01}
{F. Grosshans and P. Grangier, Phys. Rev. A \textbf{64}, 010301(R) (2001).}

\bibitem[Molmer(2006)]{Molmer06}
{K. Molmer, Phys. Rev. A \textbf{73}, 063804 (2006).}

\bibitem[Sasaki and Suzuki(2006)]{Sasaki06}
{M. Sasaki and S. Suzuki, Phys. Rev. A \textbf{73}, 043807 (2006).}

\bibitem[Leonhardt(1997)]{Leonhardt}
{U. Leonhardt, \textit{Measuring the Quantum State of Light} (Cambridge University Press, Cambridge, Uk, 1997).}

\bibitem[Biswas and Agarwal(2007)]{Agarwal07}
{A. Biswas and G. S. Agarwal, Phys. Rev. A \textbf{75}, 032104 (2007).}

\bibitem[Suzuki and al(2006)]{Suzuki06}
{S. Suzuki, K. Tsujino, F. Kannari, and M. Sasaki, Opt. Commun. \textbf{259}, 758 (2006).}

\bibitem[Wenger and al(2004)]{Wenger04}
{J. Wenger, R. Tualle-Brouri, and P. Grangier, Phys. Rev. Lett \textbf{92}, 153601 (2004).}

\bibitem[Nielsen and Molmer(2007)]{Molmer07}
{A. E. B. Nielsen and K. Molmer, Phys. Rev. A \textbf{76}, 033832 (2007).}

\bibitem[Nielsen and Molmer(2007)]{Molmer07bis}
{A. E. B. Nielsen and K. Molmer, Phys. Rev. A \textbf{75}, 043801 (2007).}

\bibitem[van Loock and Braunstein(2005)]{vanLoock05}
{S. L. Braunstein and P. van Loock, Rev. Mod. Phys. \textbf{77}, 513 (2005).}

\bibitem[Gardiner and Zoller(2004)]{Gardiner04}
{C. W. Gardiner and P. Zoller, \textit{Quantum Noise}, (Springer, New York, 2004).}


\end{thebibliography}


\end{document}